\documentclass[aps,twocolumn,reprint,superscriptaddress,showpacs]{revtex4-2}
\usepackage{amsmath,amssymb}
\usepackage{amsfonts}
\usepackage{subfigure}
\usepackage{bm}
\usepackage{graphicx}
\usepackage{appendix}
\usepackage{multirow}
\usepackage{textcomp}
\usepackage[colorlinks=true, linkcolor=blue, citecolor=blue, filecolor=magenta]{hyperref}
\usepackage{url}
\usepackage{titlesec}
\usepackage{lipsum}

\begin{document}

%\preprint{}

\title{Experimental Demonstration of Twin-Field Quantum Digital Signatures over 504 km}

\author{Chun-Hui Zhang}\thanks{These authors have contributed equally to this work.}
\author{Jing-Yang Liu}\thanks{These authors have contributed equally to this work.}
\author{Wen-Xuan Zhang}
\author{Chang Liu}
\author{Hua-Jian Ding}
\author{Xing-Yu Zhou}
\author{Jian Li}
\author{Qin Wang} \email{qinw@njupt.edu.cn}

\affiliation{Institute of Quantum Information and Technology, Nanjing University of Posts and Telecommunications, Nanjing 210003, China}
\affiliation{Key Lab of Broadband Wireless Communication and Sensor Network Technology, Ministry of Education, Nanjing University of Posts and Telecommunications, Nanjing 210003, China}
\affiliation{Telecommunication and Networks, National Engineering Research Center, Nanjing University of Posts and Telecommunications, Nanjing 210003, China}


%\date{\today}

\begin{abstract}
Digital signatures are one of the security cornerstones of the current information age. Compared with classical digital signatures based on computational complexity, quantum digital signatures (QDS) theoretically guarantee data integrity, authenticity, and non-repudiation by quantum mechanics, showing great potential for development in cryptography and thus attracting widespread attention. However, the performance of existing QDS systems are still limited in rate and distance. Here we report the first experimental demonstration of twin-field QDS (TF-QDS) using a GHz system. We achieve a maximum transmission distance of 504 km fiber spools for both single-bit and multi-bit schemes, surpassing all existing state-of-the-art QDS experiments more than 200 km.  Furthermore, by combining the one-time universal hash method, we achieve a maximum signature rate of $21.1$ times per second for a 1 Mbit file over fiber distances up to 302 km. In this work, the signature rates of both single-bit scheme and multi-bit scheme are more than two orders of magnitude higher than that of previous works at similar distance. Our work provides a new record for long-distance and high-rate QDS, representing a significant step in the development of QDS.

\end{abstract}
\maketitle

\section{INTRODUCTION}
As two cornerstones of modern cryptography, encryption and digital signatures \cite{Diffie} can protect the confidentiality, authenticity, integrity, and non-repudiation of information. Encryption ensures the confidentiality of data, while digital signatures ensure the authenticity, integrity, and non-repudiation. Compared to encrypted communication that prevents eavesdropping attacks, digital signatures actually has a much broader applications in real-world scenarios, such as email, bank transfers, e-commerce, contract agreements and blockchain. Most of these applications involving identity authentication, message authentication, and ensuring message integrity are fundamentally related to digital signatures.

Unfortunately, classical cryptography based on computational complexity \cite{RSA,ECDSA2} has been threaten by the rapid development of algorithms \cite{Shor} and computational power, especially the quantum computing \cite{QC1,QC2}. To defend these risks, quantum cryptography has emerged to provide information-theoretic security. Unlike classical methods, its security is grounded in the fundamental laws of quantum mechanics, requiring no assumptions regarding the computational capabilities of potential eavesdroppers. For example, quantum key distribution (QKD) \cite{BB84,E91} is proposed to share secure keys for encrypted communication. Similarly, quantum digital signatures (QDS) are developed to accomplish the functions of classical digital signatures. 

Since the first QDS protocol was proposed by Gottesman and Chuang in 2001 \cite{GC01},called GC01 protocol, many obstacles to practical implementations have been removed, e.g., releasing the demand of quantum memories and secure quantum channels \cite{Clarke,Arrazola1,Dunjko,Collins2014,Croal, Wallden,Yin2016,Amiri,Donaldson}. Particularly, in Refs. \cite{Wallden,Yin2016,Amiri}, utilizing QKD components and process as key generation protocol (KGP) for signing message was proposed, and Ref. \cite{Amiri} presented a general security analysis with decoy method \cite{WXB2005,Lo}, making QDS more practical and easier to realize. Thereafter, studies focus primarily on practical performance of QDS, such as high rates \cite{Collins2017,AnXB,Richter}, long distances \cite{Yin2017,Ding}, practical security \cite{MDIQDS,Roberts,Zhang}, field trials and networks \cite{Yin2017field,Chapman,Pelet}. To further improve QDS performance, we proposed the twin-field (TF)-QDS \cite{TFQDS}, significantly extending the signature distance. These studies mentioned so far still follow the idea of GC01 protocol that requires long states or keys to sign one bit document once, which is called single-bit-type QDS. To improve the efficiency, Yin \textit{et al.} proposed a multi-bit QDS scheme named one-time universal hash (OTUH)-QDS protocol \cite{Yin2023,Li2023},  substantially boosting the signature rates. It has been accomplished in chip-integrated QDS system \cite{ChipQDS} and shown the application potential in quantum e-commerce \cite{Cao2024}.

To date, many QKD experiments have achieved transmission distance exceeding 500 km \cite{Pittaluga600, Chen511,Chen658,Wang830,Zhou615,Zhou508AMDI} even 1000km \cite{SatelliteQKD,TFQKD1000} in free space or fiber, predominantly utilizing the TF-QKD protocol \cite{TFQKD,SNSTF,PMQKD,CuiC,Curty}. In contrast, the maximum distance for QDS has remained limited to 280 km \cite{Ding}, and an experimental demonstration of TF-QDS is still absent. This disparity significantly constrains the practical utility of QDS. In this paper, we report the first experimental demonstration of a three-party TF-QDS system operating at a GHz clock rate.  For the TF-QDS protocol, we accomplish the proof-of-principle demonstration of both the single-bit scheme \cite{TFQDS} and multi-bit scheme \cite{Yin2023}, where a 3-intensity sending-or-not-sending (SNS) protocol with actively-odd-parity-pairing (AOPP) method \cite{AOPP,AOPP2} is used during the key generation process. 
Our experiments reached a landmark distance of 504 km, a significant extension over prior works. Notably, at 302 km, the system yields a signature rate of $21.1$ tps (for 1 Mbit documents), showcasing both high efficiency and exceptional reach. This study represents a critical milestone in the transition of QDS from proof-of-concept to high-performance practical application.

\section{TF-QDS PROTOCOL}
\begin{figure}[]
	\centering
	\includegraphics[width=1.0\linewidth]{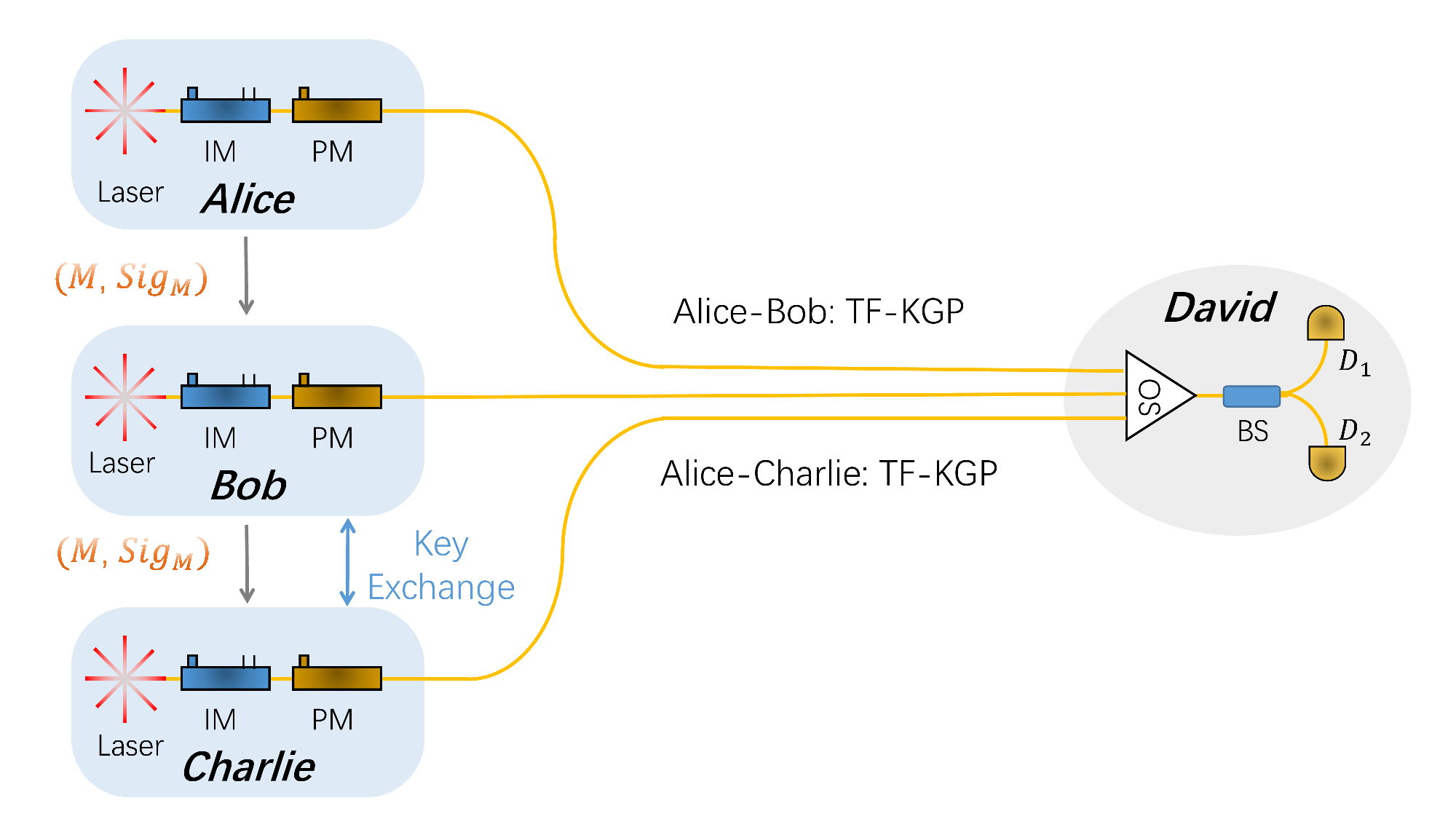}
	\caption{Brief framework of TF-QDS. In distribution stage, Alice-Bob and Alice-Charlie independently perform TF-KGP with an optical switch (OS), where the weak coherent states (WCS) of Laser at each side are modulated by intensity modulators (IM) and phase modulators (PM), and then sent to an untrusted party David for measurement by a beam splitter (BS) and two detectors ($D_1$ and $D_2$).  In messaging stage, Alice transmits the message and signature $(M,Sig_M)$ to Bob, and Bob then forwards them to Charlie. Finally, Bob and Charlie verify  the signature. In the process of TF-QDS, there are a step of key exchange between Bob and Charlie. }
	\label{fig1}
\end{figure}

In this section, we firstly present a general architectural framework of three-user TF-QDS that incorporates the single-bit scheme \cite{TFQDS} and multi-bit scheme \cite{Li2023}, as shown in Fig. \ref{fig1}.  Here, Alice is the signer of message, and Bob and Charlie are the verifiers. It begins with Alice performing TF-KGP with both Bob and Charlie independently through an untrusted party David, enabling all three parties to obtain the keys for subsequent signature and verification. Next, Alice signs the message and transmits the message and signature $(M,Sig_M)$ to Bob, who then transmits it to Charlie. Additionally, during the process, Bob and Charlie need to exchange specific verification keys to each other. Finally, both verifiers check the signature and make an acceptance decision. 

No matter single-bit scheme or multi-bit scheme, there are two stages, i.e., distribution stage and messaging stage. Since the two schemes differ in their procedural sequences and the precise content of transmitted information, we introduce the detailed workflow as follows.

\subsection{Distribution stage}
Firstly, Alice needs to perform TF-KGP with Bob and Charlie respectively, and we adopt the SNS-TF protocol with the AOPP method \cite{AOPP,AOPP2} as the KGP. Alice-Bob and Alice-Charlie individually generate $N$ photon pulses and code them using intensity modulator (IM) and phase modulator (PM). Each pulse is randomly chosen as the $X$ or $Z$ window during this process. In the $X$ window, each side randomly prepares and sends out weak coherent states, vacuum state $o$ and decoy state $v$. In the $Z$ window, a signal state $u$ is sent with a probability $p_s$, and nothing $(0)$ is sent with probability $1-p_s$. The prepared pulses are sent to the measurement station which is controlled by David, who is assumed to measure the interference results of the incoming pulse pair and announce the measurement results to the two parties of TF-KGP. If one and only one detector click is announced, the time window is regarded as an effective window, and the corresponding event is an effective event. After receiving all the measurement results, they sift the effective results of both using $Z$ window as raw key bits and then publish partial bits for error test.

\emph{\textbf{Single-bit scheme:}} 
The raw keys are as key pool, then Bob and Charlie randomly select half of $\mathcal{L} $-length keys in key pool to exchange via authenticated classical channel, as in Fig. \ref{fig1}. 

\emph{\textbf{Multi-bit scheme:}}
Alice–Bob (Alice–Charlie) perform the error correction on all raw keys to obtain identical keys as key pool.

\subsection{Messaging stage}
\emph{\textbf{Single-bit scheme:}}
In signing single-bit message $m$ (0 or 1), Alice–Bob (Alice–Charlie) choose corresponding bit string $A_B^m$ and $S_B^m$ ($A_C^m$ and $S_C^m$) from the key pool, where $A_B^m$ ($A_C^m$) are the keys of Alice generated from the TF-KGP of Alice–Bob (Alice–Charlie). $S_B^m$ ($S_C^m$) are the keys including on half from TF-KGP of Alice–Bob (Alice–Charlie) and another half from Charlie's (Bob's) exchange. Then, Alice transmits the signature ($m$, $S_m$) to Bob, where $S_m = (A_B^m, A_C^m)$. Bob verifies the signature by comparing $S_B^m$ with the corresponding portion of $S_m$, recording the number of mismatches. If the mismatch counts are below $s_a\mathcal{L} /2$, the signature is accepted; otherwise, it is rejected. Here, $s_a$ represents an acceptance threshold. Subsequently, Bob forwards the signature ($m$, $S_m$) to Charlie. Charlie performs an analogous verification process by comparing $S_C^m$ with $S_m$. The signature is accepted only if the mismatch counts are below $s_v\mathcal{L} /2$, where $s_v$ is an another verification threshold satisfying $s_a < s_v < 0.5$. Otherwise, the signature is rejected. For long message $M$, they repeat single-bit signing process of $M$ times.

\emph{\textbf{Multi-bit scheme:}} 
For high-rate signatures, the OTUH method is used to sign the multi-bit long message $M$ \cite{Li2023}.   Alice employs a generalized division hash function $H_p$, constructed from a locally generated $\mathcal{L}$-bit random number $p_a$ to compute an $\mathcal{L}$-bit hash value $h = H_p(M)$ of the long message $M$. Next, Alice encrypts the hash value $h$ with the bit string $Y_a$ and the random number $p_a$ with the bit string $X_a$, obtaining the signature $S = h \oplus Y_a,$ and $P = p_a \oplus X_a$. Finally, Alice transmits the set \{$S$, $P$, $M$\} to Bob through an authenticated classical channel to complete the signature process.
Bob transmits the set \{$S$, $P$, $M$\} and his bit strings \{$X_b$, $Y_b$\} to Charlie after receiving the message from Alice. Similarly, Charlie transmits his bit strings \{$X_c$, $Y_c$\} to Bob upon receipt of the message from Bob. This corresponds to the signature transmission and key exchange process shown in Fig. \ref{fig1}. Critically, all transmissions occur through authenticated channels. Using the XOR operation, Bob (Charlie) can generate two bit strings $K_{X_b} = X_b \oplus X_c, K_{Y_b} = Y_b \oplus Y_c$ ($K_{X_c} = X_b \oplus X_c, K_{Y_c} = Y_b \oplus Y_c$), then obtains the expected digest $h_b = S \oplus K_{Y_b}$ ($h_c = S \oplus K_{Y_c}$) and $p_b = P \oplus K_{X_b}$ ($p_c = P \oplus K_{X_c}$).
Finally, Bob (Charlie) uses $p_b$ ($p_c$) to construct the generalized division hash function $H_{p_b}$ ($H_{p_c}$) and computes an actual digest $H_b = H_{p_b}(M)$ ($H_c = H_{p_c}(M)$). If $H_b = h_b$ and  $H_c = h_c$ , they accept the signature; otherwise, they reject it. 

With the key pool of length $n_{pool}$ and the $\mathcal{L} $-bit signature, we derive the signature rate
\begin{align}
	R &= \frac{n_{pool}}{2\mathcal{L} }.
\end{align}
The detailed TF-KGP introduction and parameter estimation are presented in Appendix \ref{APP:KGP}, and the security analysis of single-bit scheme and multi-bit scheme are given in Appendix \ref{APP:security}.

\section{EXPERIMENTAL SYSTEM AND RESULTS}
\begin{figure*}[htbp]
 	 \centering
	 \includegraphics[width=\textwidth]{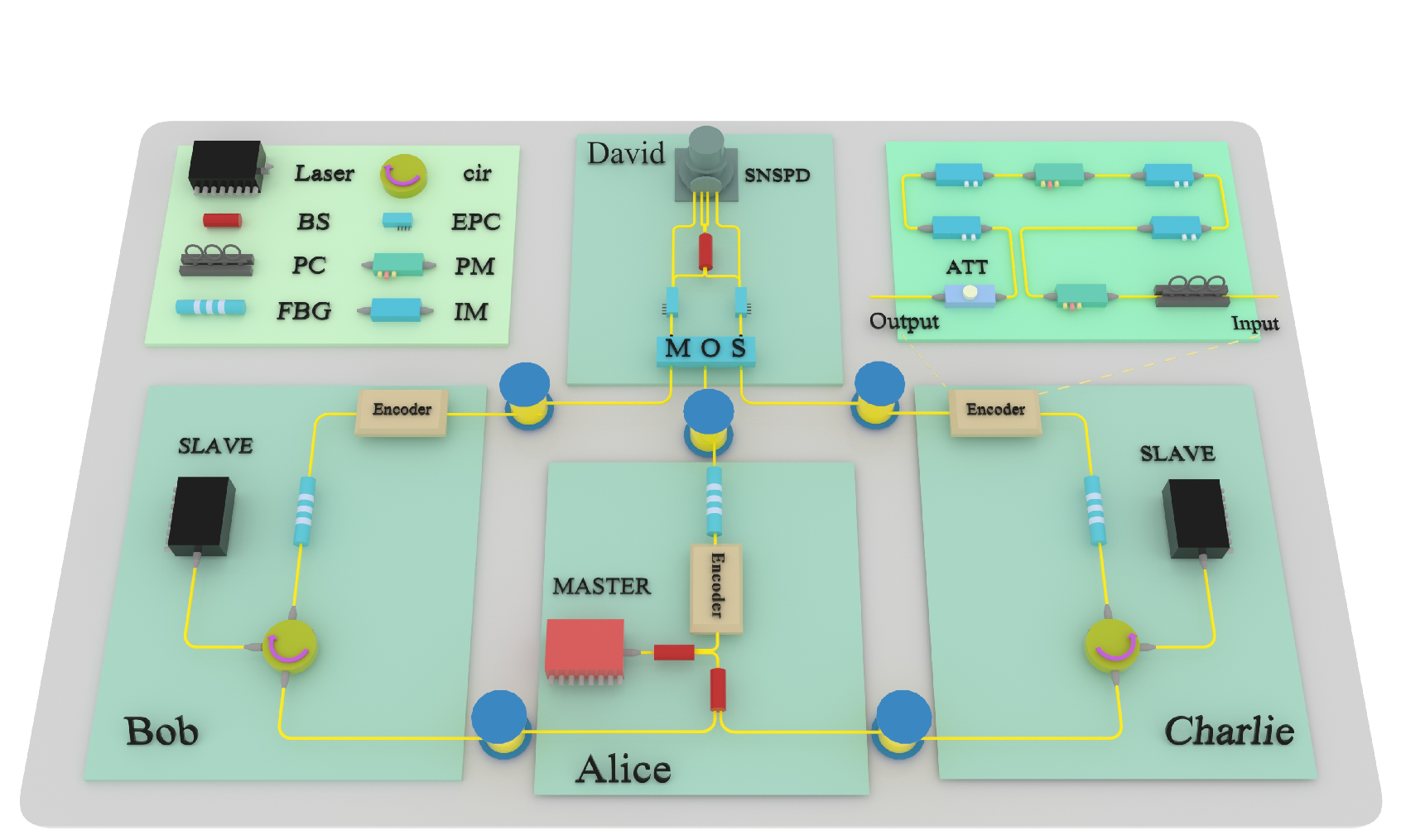}%
	 \caption{Experimental setup of TF-QDS. In the system, the light of Alice's master laser is split into two beams, in which one beam is for encoding, and the other is further split and respectively transmitted to Bob and Charlie. The light passes through the circulator (Cir) and injects the slave laser for seeding. By seed injection, Bob or Charlie separately locks the laser frequency with Alice's laser. Next, their light are filtered by a fiber Bragg grating (FBG), and then modulated by a encoder, which consists of polarization controller (PC), phase modulators (PMs), intensity modulators (IMs) and attenuator (ATT).  The IMs are used to set the intensities of reference and signal pulses, and the PMs are used to encode the phase of pulses. The pulses are then attenuated by an ATT and sent out via fiber spools to David. At David’s measurement station, a multi-port optical switch (MOS) is used to choose which two parties' (Alice-Bob or Alice-Charlie) pulses for further measurement. Then the pulses pass through polarization  compensation module, including a electronic polarization controller (EPC) and a polarization beam splitter (PBS). Finally, the pulses interfere at a beam splitter (BS) and are detected by superconducting nanowire single-photon detectors (SNSPDs). }
	 \label{fig2}
\end{figure*}

The experimental setup used in our work is shown in Fig. \ref{fig2}. It consists of four main modules: the senders Alice, Bob, Charlie, and the intermediate measurement node David. Here, Alice, Bob and Charlie each prepares three intensities ($u, v, 0$) to perform the SNS-TF-KGP process. Each pulse is randomly chosen as the $X$ or $Z$ window, where In the $X$ window, the vacuum state $o$ and decoy weak coherent states $v$ are randomly prepared with probability $p_o$ and $p_v$. In the Z window, the signal state $u$ is sent with probability $p_s$, and vacuum is sent with probability $1-p_s$. 

In TF-KGP, we should ensure the stability of the frequency difference between Alice's and Bob's (Charlie's) independent lasers. Here, we lock the frequencies of the lasers on each side by using an optical injection locking (OIL) technique \cite{Fang2020,Du2024} to distribute a common optical phase reference. Specifically, as shown in Fig. \ref{fig2}, Alice's continuous-wave (CW) laser with central wavelength at 1550.12 nm and linewidth of 0.1 kHz serves as the master laser, whose output is split by a 50:50 beam splitter (BS) into two beams. One beam is sent to an encoder module for state preparation, and another beam is further split into two seeding beams and delivered to Bob and Charlie. At Bob’s and Charlie's stations, a distributed-feedback (DFB) laser serves as the slave laser and is optically injected after the corresponding seed light passing through a circulator (CIR). Under optical injection locking, the slave lasers are forced to emit optical fields that inherit the frequency and phase of the master laser. As a result, Bob’s and Charlie's generated optical pulses originate from a common phase reference.

\begin{table}
	\caption{System parameters in this experiment. $\eta_d$ and $P_{d}$ are the efficiency and dark count rate of our SNSPDs; $M$ is number of phase slice used in our system; $e_d$ is the misalignment error rate in TF-KGP; $\alpha$ is the attenuation coefficient of fiber; $\epsilon$ is the security level of TF-QDS system.}
	\begin{tabular}{c c c c c c}
		\hline \hline
		 $\eta_d$  &  $P_{d}$   & $M$ & $e_d$ &  $\alpha$ (dB/km) & $\epsilon$  \\ \hline
		  69\% \quad &  $8\times10^{-11}$ \quad & 16 \quad  & \quad 1.8\% & \quad 0.19 & \quad $10^{-10}$   \\
		\hline \hline
	\end{tabular}
	\label{tab:parameters}
\end{table}

After the wavelengths and phases of the lasers at Alice's and Bob's ends are locked using OIL, the next step is encoding. The injection-locked slave lasers are modulated by radio-frequency signals generated by an arbitrary waveform generator and amplified before driving the lasers, producing optical pulses at a clock rate of 1.25 GHz. The generated pulses of slave lasers pass through a 10 GHz fiber Bragg to suppress unwanted spectral components due to the amplified spontaneous emission noise.
Then, the optical pulses enter the encoder module, which consists of polarization controller (PC), intensity modulators (IMs), phase modulators (PMs) and attenuator (ATT). The first PM are used for phase randomization, and the second PM is used prepare the phases of coherent states. The first two IMs are used for preparing reference light and quantum light, while later two IMs are used to modulate the intensity $u, v, o$ of the coherent states. 

It is important to accurately estimate and compensate for the relative phase drift from the two long fiber channels. Here, we adopt a time-division multiplexed scanning calibration scheme to perform phase post-estimation on the global phase drift \cite{Wang830,TFQKD1000}. Specifically, we estimate the relative phase between Alice and Bob with the strong reference light, then corrected the relative phase in post-processing. Here, $\mathrm{IM}_1$ and $\mathrm{IM}_2$ chop the pulse train into the reference light and quantum light parts. The brighter reference light is responsible for generating error signals for phase drift and also serves as a guide to the outcome of the compensation. During each 40 $\mu$s time period, we send  pulses at a system frequency of 1.25 GHz with a period 800 ps. In the first 7.6 $\mu$s of 40 $\mu$s, we modulate reference pulses into two different phases (0, $\pi/2$) in turn, while Bob maintained the phase at 0. Following the strong reference pulses, a duration of 0.4 $\mu$s is designated for the recovery period for the SNSPDs, when both Alice and Bob modulate the pulses to vacuum. The pulses of left 32 $\mu$s are randomly prepared in 16 phase slices. Hence the duty cycle of our system is 80\%, and the effective frequency is 1 GHz. Finally, the pulses are attenuated to the desired levels using an electrical variable optical attenuator (ATT) before transmission to Charlie through low-loss fibers with an average attenuation of less than 0.19 dB/km.

After passing through the corresponding quantum channels, the pulses from Alice, Bob and Charlie first pass through a multi-port optical switch (MOS), where which two parties (Alice-Bob or Alice-Charlie) are decided to perform TF-KGP. After MOS, the pulses are aligned to the same polarization by a polarization compensation module, which includes electrical polarization controllers (EPC) and polarization beam splitters (PBS). Then, a 50:50 BS performs a single photon interference on the incoming pulses. Finally, the twin field interfering results are detected by two SNSPDs, and recorded by a time tagger. To achieve the necessary high efficiency for pulse detection, we place high performance superconducting nanowire single-photon detectors (SNSPD) at David, whose average detection efficiency and dark count rate are 69\% and less than 0.1 Hz.

With above experimental setup, we performed a series of experiments on both multibit TF-QDS and single-bit TF-QDS over transmission distances of 302 km and 504 km, with each SNS-TF-KGP involving the transmission of a total of $10^{14}$ pulses. Specifically, signatures are generated for a 1 Mbit message using the multi-bit TF-QDS scheme. The security level $\epsilon$ of our TF-QDS system in both multi-bit and single-bit scheme is set as $10^{-10}$.  Table \ref{tab:parameters} shows the parameters used in our experiments and simulations and we performed global optimization of parameters to maximize the signature rate in the simulations. And the detailed observed values and estimated parameters are presented in Appendix \ref{APP:data}.

\begin{figure}[]
	\centering
	\includegraphics[width=1.0\linewidth]{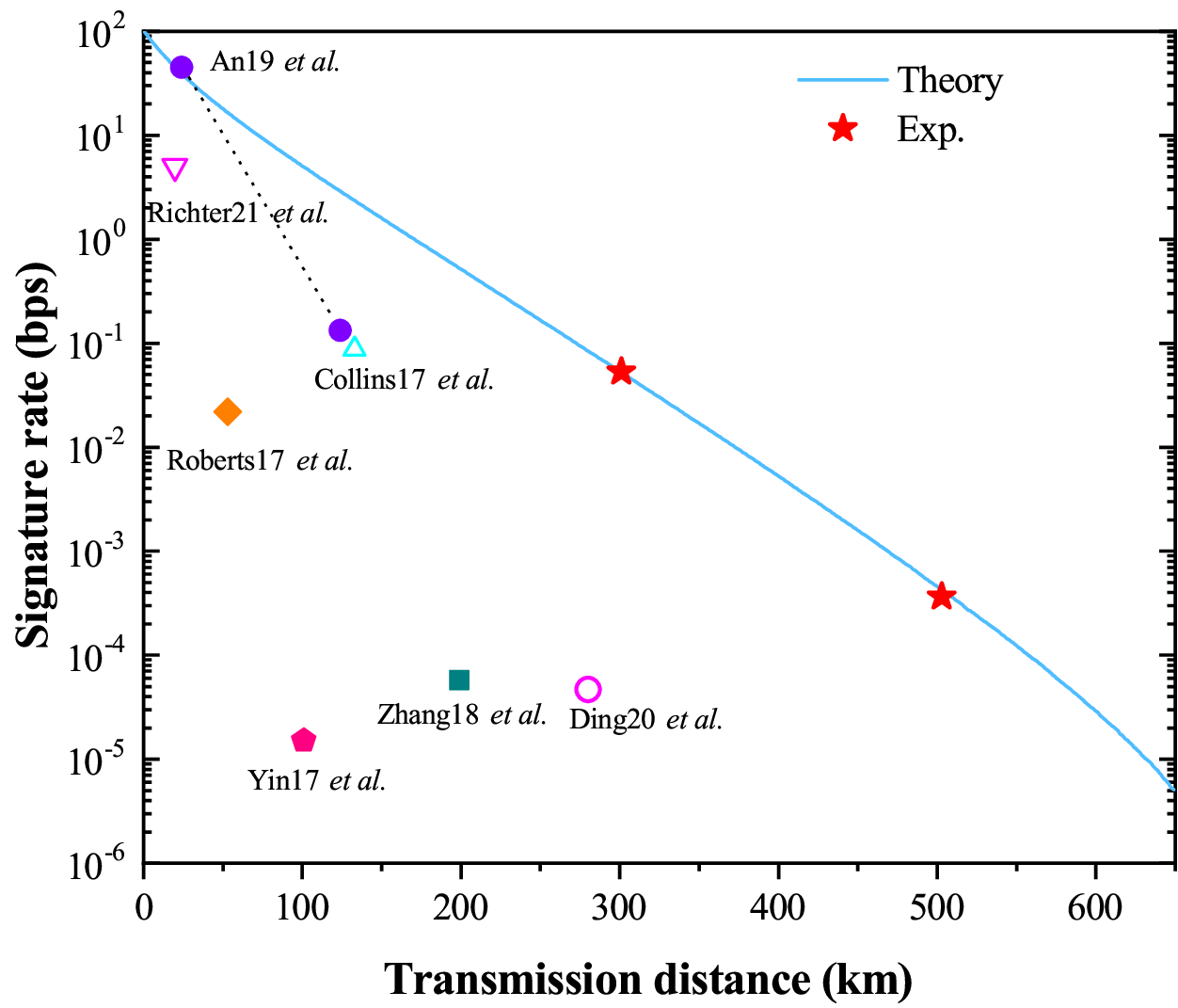}
	\caption{Signature rates (bps) of different works based on single-bit scheme. The solid curve and pentagram dots correspond to the simulation and experimental signature rates. Other works are An \emph{et al.} \cite{AnXB}, Richter \emph{et al.} \cite{Richter},  Collins \emph{et al.} \cite{Collins2017}, Roberts \emph{et al.} \cite{Roberts}, Yin17 \emph{et al.} \cite{Yin2017}, Zhang \emph{et al.} \cite{Zhang}, Ding \emph{et al.} \cite{Ding}.}
	\label{fig3}
\end{figure}

\begin{figure}
	\centering
	\includegraphics[width=1.0\linewidth]{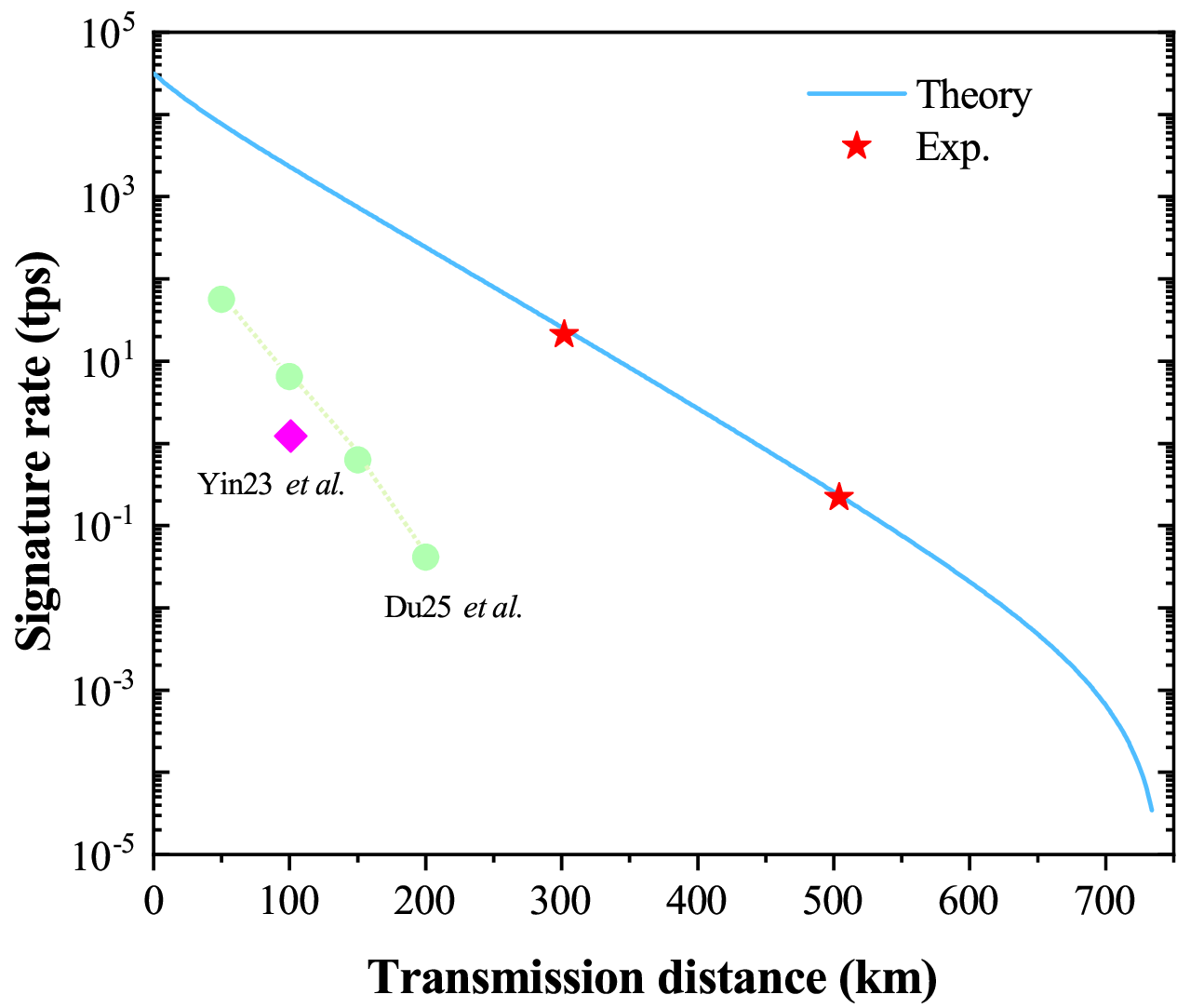}
	\caption{Signature rates (tps) of different works using multi-bit scheme. Here the rate is the times of signing $1$ Mbits message per second. The solid curve and pentagram dots correspond to the simulation and experimental signature rates in our experiment. Other works are Yin23 \emph{et al.} \cite{Yin2023}, Du \emph{et al.} \cite{ChipQDS}.}
	\label{fig4}
\end{figure}

We achieve signature rates of $0.0532$ bps, $3.65\times10^{-4}$ bps at transmission distances of 302 km, and 504 km, respectively. Fig. \ref{fig3} presents our experimental results (red star) and theoretical simulation (solid line) of the single-bit TF-QDS scheme, where the experimental results demonstrate agreement with the simulation data. To illustrate the advantage of our work, we compare it with the signature rates of other works in recent years \cite{Collins2017,AnXB,Richter,Yin2017,Ding,Roberts,Zhang}, as shown in Fig. \ref{fig3} and Table \ref{tab:comparion}. The signature rate of our system at $302$ km is $0.0532$ bps, which is almost three orders of magnitude higher than that of Ding \emph{et al}'s work \cite{Ding} at $280$ km. Besides, we achieve the longest distance record of 504 km, which is more than $200$ km longer than previous works.

For the multi-bit TF-QDS scheme, we also conducted experiments at transmission distances of 302 km and 504 km. In multi-bit scheme, the signature rate is defined as time per second (tps) of signing 1 Mbit message. As shown in Fig. \ref{fig4}, we demonstrate the signature rates for $21.1$ tps and $0.219$ tps at 302 km and 504 km, respectively. For detailed quantitative comparison, we list the parameters and results several representative single-bit and multi-bit signature experiments, as summarized in Table \ref{tab:comparion}. Compared to other multi-bit QDS works scheme based on OTUH \cite{Yin2023,ChipQDS}, the signature rate of our work is significantly higher, and the transmission distance is 300 km longer than their works. Specifically, the signature rate of this work at $302$ km is 500 times higher than that of Du \emph{et al}'s work \cite{Ding} at $200$ km. This significant advantage primarily stems from the TF-QDS protocol and our high-speed experimental system of GHz.

\begin{table*}
	\centering
	\caption{Comparisons of signature rate between different work on QDS.}
	\begin{tabular}{ccccccc}  \hline \hline
		Reference   \quad & KGP & Clock rate & Maximal  distance & Message size & Security level & Signature rate \\ \hline
		An \emph{et al.} \cite{AnXB}  \quad & BB84 & $1$ GHz & $125$ km & 1 bit & $10^{-5}$ & $0.133$ bps \\
		Richter \emph{et al.} \cite{Richter} \quad &  CV  & $1$ GHz & $20$ km & 1 bit & $10^{-4}$ & $5$ bps \\
		Collins \emph{et al.} \cite{Collins2017} \quad &  DPS  &  $1$ GHz &  $134$ km &  1 bit  &  $10^{-4}$ & $0.088$ bps\\
		Roberts \emph{et al.} \cite{Roberts}   \quad &  MDI &  $1$ GHz &  $50$ km &  1 bit &  $10^{-10}$ & $0.022$ bps \\
		Yin17 \emph{et al.} \cite{Yin2017}   \quad &  SARG04 &  $75$ MHz &  $102$ km &  1 bit &  $10^{-9}$ & $1.5\times10^{-5}$ bps \\
		Zhang \emph{et al.} \cite{Zhang}   \quad &  BB84 &  $76$ MHz &  $200$ km &  1 bit &  $10^{-4}$ & $5.75\times10^{-5}$ bps \\
		Ding \emph{et al.} \cite{Ding}   \quad &  BB84 &  $50$ MHz  &  $280$ km &  1 bit &  $10^{-5}$ & $4.67\times10^{-5}$  bps \\
		Yin23 \emph{et al.} \cite{Yin2023}   \quad &  BB84 &  $200$ MHz &  $101$ km &  $10^6$ bits &  $10^{-32}$ & $1.22$ tps \\
		Du \emph{et al.} \cite{ChipQDS}   \quad &  BB84 &  $50$ MHz &  $200$ km &  $10^6$ bits &  $4.72\times10^{-8}$ & $0.0414$ tps \\
		\multirow{2} *  { \textbf{This work} } \quad  &  \multirow{2} * {\textbf{TF}} &  \multirow{2} * {$\mathbf{1.25}$ \textbf{GHz}} &  \multirow{2} * {$\mathbf{504}$ \textbf{km}} & \multirow{2} * {$\mathbf{10^6}$ \textbf{bits}}  &  \multirow{2} * {$\mathbf{10^{-10}}$} & $\mathbf{21.1}$ \textbf{tps} $\mathbf{@302}$ \textbf{km} \\ & & & & &  & $\mathbf{0.219}$ \textbf{tps} $\mathbf{@504}$ \textbf{km} \\ \hline\hline
	\end{tabular}
	\label{tab:comparion}
\end{table*}

\section{CONCLUSION}
In summary, we have successfully demonstrated the first experimental implementation of TF-QDS utilizing a high-speed GHz-clock-rate system. By overcoming the fundamental limitations of previous protocols, we achieved a record-breaking transmission distance of 504 km, the longest reported to date for any QDS system. Our work showcases the versatility of the twin-field architecture by successfully executing both single-bit and multi-bit TF-QDS schemes. And no matter in which scheme, our work has more than $200$ km improvement in transmission distance and two orders of magnitude in signature rates enhancement at similar distances compared to all existing works. Notably, this implementation inherently provides measurement-device-independent security, offering a robust defense against all detector side-channel attacks. These results represent a substantial leap in performance compared to prior state-of-the-art QDS experiments, marking a pivotal milestone toward the deployment of high-performance, long-haul quantum digital signature networks.

\begin{acknowledgments}
We acknowledge the support of Industrial Prospect and Key Core Technology Projects of Jiangsu provincial key R\&D Program (BE2022071), National Natural Science Foundation of China (62471248, 62401287, 12074194) .
\end{acknowledgments}

\appendix
\section{KEY GENERATION PROTOCOL}\label{APP:KGP}
In this section, we introduce the TF-KGP used in our experiment, which is the 3-intensity SNS-TF protocol with AOPP \cite{AOPP,AOPP2} without error correction and privacy amplification.

The prepared phase-randomized coherent state can be expressed as
\begin{align}\label{Coherent}
	\left| {\sqrt x_U {e^{i\theta_U }}} \right\rangle  = \sum\nolimits_{n = 0}^\infty  {\frac{{{e^{ - x_U/2}}{{(\sqrt x_U {e^{i\theta_U }})}^n}}}{{{\sqrt {n!} }}}\left| n \right\rangle } , 
\end{align}
where $U \in \{A, B, C\}$ denotes which user (Alice, Bob, or Charlie). $x_U$ and $\theta_U$ represent the intensity and phase of coherent state, respectively. Here, $x_U \in \{o, v, u\}$ and $\theta_U$ are random in $[0,2\pi)$. The users randomly choose a vacuum state ($o$), decoy states ($v$), and $Z$ windows with probabilities $p_o, p_v, p_Z$, respectively, where $p_o+p_v+p_Z=1$. When a $Z$ window is chosen, they send a signal state $u$ with probability $p_s$, and send nothing with $1-p_s$.

When receiving the pulses from two users, such as Alice and Bob, David performs measurements and announces the results. During the measurement process, if only one detector clicks, David announces a successful event, recorded as a one-detector heralded event, and announces which detector ($D_1$ or $D_2$) clicks. When the measurement process is complete and results have been announced,  publicly disclose which window was used for each pulse pair. When Alice and Bob both use $Z$ windows, called $Z$ basis, the phase and sending-or-not-sending operation should be never disclosed. Otherwise, they disclose the state intensity, and if they both send decoy states $v$, the phases should be also disclosed. Furthermore, we need to post-select the effective events of both Alice and Bob sending coherent states $v$, where the phases of two states $v$ satisfy the following criterion:
\begin{align}\label{PhaseSlice}
	\left|\theta_{A}-\theta_{B (C)} -\psi_{\mathrm{AB (AC)}}-k\pi\right| \leq \frac{\Delta}{2}, 
\end{align}
which is the case called $X$ basis.
In Eq. (\ref{PhaseSlice}), $\psi_{\mathrm{AB (AC)}}$ is the difference of global phases between Alice-David's link and Bob-David's link (Charlie-David's link), which results in the optical misalignment error ($e_d$); $k=0,1$ corresponds to in-phase or anti-phase of ${\theta _A}$ and ${\theta _B}$ (${\theta _C}$); $\Delta=\frac{{2\pi }}{M}$ represents the size of each slice, and $M$ refers to the total number of phase slices pre-chosen by Alice and Bob.

For the one-detector heralded events on $Z$ basis, Alice (Bob) denotes it as bit 0 if she (he) sends nothing (state $u$) and as bit 1 if she (he) sends state $u$ (nothing). For the effective events of $X$ basis, a right click is the $D_1$ ($D_2$) detector clicking when $k=0$ ($k=1$), and a wrong click is the $D_2$ ($D_1$) detector clicking when $k=0$ ($k=1$).
The data on $Z$ basis are used for the error test and signature, and finally Alice and Bob form an $n_Z$-length key string $Z_s$ and $Z^\prime_s$, respectively. Other data are used to estimate single-photon contributions, i.e. the counts and error rates of the single-photon components ($\underline{n}_{1}, \overline{e}_{ph}$) on $Z$ basis.

Firstly, we estimate the lower bound of single-photon counts and upper bound of single-photon error counts on $X$ basis considering finite-size effect. From the observed values, we know the counts of one-detector heralded events with various intensity combinations ($n_{ab}$, $a,b \in \{o,v,u\}$), and the counts of error clicks in $X$ basis $m_{vv}$. With these observed values, we obtain
\begin{align}\label{n1X}
	n_{X,1}  = &\frac{{{\tau _{X,1}}}}{{2uv(u - v)}}\Biggl[ \frac{{{u^2}{e^v}({n^-_{ov}}+{n^-_{vo}}) }}{{{p_o p_v}}} - \notag \\
	 &\frac{{{v^2}{e^u}({n^+_{ou}}+{n^+_{uo}}) }}{{{p_op_Zp_s}}} - \frac{{2({u^2} - {v^2}){n^+_{oo}}}}{{{p^2_o}}} \Biggr],
\end{align}
\begin{align}\label{eph}
	 e_{ph}  =  {\frac{{{m^+_{vv}-\frac{{e^{-2v} p^2_v}}{{Mp_o^2}}n^-_{oo}}}}{{ n_{X,1}}} } ,
\end{align}
where $\tau _{X,1}$ is the probability of single-photon components in interfering phase slice on $X$ basis, which is ${\tau _{X,1}} = 4p_v^2 ve^{-2v}/M$. 
The $x^-$ and $x^+$ in Eqs. (\ref{n1X}) and (\ref{eph}) are the observed values when considering the statistical fluctuations by the Chernoff bound \cite{Chernoff}
\begin{align}
	x^-=\phi^L(x) &= \frac{x}{1 + \delta_1(x)}, \label{chernoff_begin}\\
	x^+=\phi^U(x) &= \frac{x}{1 - \delta_2(x)}, 
\end{align}
where we can obtain the values of $\delta_1(x)$ and $\delta_2(x)$ by solving the following equations:
\begin{align}
	\Biggl( \frac{e^{\delta_1}}{(1+\delta_1)^{1+\delta_1}} \Biggr)^{x/(1+\delta_1)} &= \frac{\epsilon_p}{2},  \\
	\Biggl( \frac{e^{-\delta_2}}{(1-\delta_2)^{1-\delta_2}} \Biggr)^{x/(1-\delta_2)} &= \frac{\epsilon_p}{2}, \label{chernoff_end}
\end{align}
where $\epsilon_p$ is the failure probability.

Next, we can similarly estimate the lower bound of single-photon counts in untagged bits of  $Z$ basis by
\begin{align}\label{nu0}
	n_{u0}  = \frac{{{\tau _{Z,1}}}}{{uv(u - v)}}\Biggl[ \frac{{{u^2}{e^v}{n^-_{ov}} }}{{{p_o p_v}}} - 
	 &\frac{{{v^2}{e^u}{n^+_{ou}}}}{{{p_op_Zp_s}}} - \frac{{({u^2} - {v^2}){n^+_{oo}}}}{{{p^2_o}}} \Biggr],
\end{align}
\begin{align}\label{nu1}
	n_{u1}  = \frac{{{\tau _{Z,1}}}}{{uv(u - v)}}\Biggl[ \frac{{{u^2}{e^v}{n^-_{vo}} }}{{{p_o p_v}}} - 
	 &\frac{{{v^2}{e^u}{n^+_{uo}}}}{{{p_op_Zp_s}}} - \frac{{({u^2} - {v^2}){n^+_{oo}}}}{{{p^2_o}}} \Biggr],
\end{align}
where ${\tau _{Z,1}} = p_Z^2 p_s(1-p_s)ue^{-u}$; $n_{u0}$ is the number of the untagged bits 0 and $n_{u1}$ is the number of the untagged bits 1.

In the AOPP method, the bit 0 and bit 1 would be actively paired by Alice-Bob (Alice-Charlie) at corresponding keys position. After AOPP, the number and error rate of remaining keys are $n^\prime_{Z}$ and $E^\prime_{Z}$, and $n_{pool}=n^\prime_{Z}$. By taking the estimated values before AOPP, we can calculate the number of single-photon untagged bits $n_{1}^{\prime}$ and the single-photon phase-flip error rate  $e_{ph}^{\prime}$ after AOPP by the method proposed in Ref. \cite{AOPP,AOPP2}. We have the related formulas as follows:

\begin{align}
	&h=\frac{n_g}{2n_{odd}}, \\
	&n_{u1}^L=\phi^L(n_{u1} h), \\
	&n_{u0}^L=\phi^L(n_{u0} h),\\
	&n_1^L=n_{u1}^L+n_{u0}^L,\\
	&n_1^r=\phi^L\left(\frac{(n_1^L)^2}{2hn_{Z}}\right), \\
	&   n^\prime_{u1}=2n_1^r\left(\frac{n_{u1}^L}{n_1^L}-\sqrt{-\frac{\ln\varepsilon}{2n_1^r}}\right),
\end{align}
\begin{align}
	&	n^\prime_{u0}=2n_1^r\left(\frac{n_{u0}^L}{n_1^L}-\sqrt{-\frac{\ln\varepsilon}{2n_1^r}}\right),\\
	&	n_{min}=\min(n^\prime_{u1},n^\prime_{u0}),\\
	&		r=\frac{n_1^L}{n_1^L-2n_1^r}\ln\frac{3(n_1^L-2n_1^r)^2}{\varepsilon},\\
	&		e_T=\frac{\phi^U(2n_1^r e_{ph}^U)}{2n_1^r-r},\\
	&	M_S^U=\phi^U[(n_1^r-r)e_T(1-e_T)]+r,
\end{align}
where $n_Z$ is number of raw keys that Alice and Bob get in the experiment; $n_{odd}$ is the number of pairs with odd parity if Bob randomly groups all his raw key bits two by two, and $n_g$ and $n_{odd}$ are observed values; $\varepsilon$ is the failure probability of parameter estimation. Then the lower bound of $n_{1}^{\prime}$ and the upper bound of $e_{ph}^{\prime}$ by
\begin{align}
  &n^\prime_{1}=2\phi^L\left(n_{min}(1-\frac{n_{\min}}{2n_1^r})\right),\\
  &e_{ph}^\prime=\frac{2M_S^U}{	n^\prime_{1}}.
\end{align}

To maximize the signature rate, we need to find the minimum length $\mathcal{L}$ of signature that satisfies the security level $\epsilon$. In single-bit scheme, the truly secure keys are the half of keys (length $\mathcal{L}/2$) kept by Bob and Charlie, which are not exchanged. Hence, we need further estimate single-photon counts and phase error rate in the $\mathcal{L}/2$-bit string, which can be estimated with the calculated values of the parameters above by using the random sampling without replacement\cite{Li2023,Yin2020}:
\begin{align}\label{n1Ls}
	n^{1}_{\mathcal{L}} \geq \frac{\mathcal{L}}{2}\left[\frac{n^\prime_{1}}{n^\prime_{Z}} - \gamma^U\left(\frac{\mathcal{L}}{2}, n^\prime_{Z} - \frac{\mathcal{L}}{2}, \frac{n^\prime_{1}}{n^\prime_{Z}}, \epsilon_p\right)\right],  
\end{align}
\begin{align}\label{ephLs}
	e^{ph}_{\mathcal{L}} \leq e_{ph}^\prime + \gamma^U(n^{1}_{\mathcal{L}}, n^\prime_{1} - n^{1}_{\mathcal{L}}, e_{ph}^\prime, \epsilon_p).
\end{align}
Therefore, in single-bit scheme, the min-entropy $\mathcal{H}_\mathcal{L} $ can be estimated in the distribution stage as
\begin{align}
	\mathcal{H}_\mathcal{L}  \leq n^{1}_{\mathcal{L}} [1 - H(e^{ph}_{\mathcal{L}})] -\frac{\mathcal{L}}{2} H(E^\prime_Z).\label{Hns}
\end{align}

In multi-bit scheme, we need estimate single-photon counts and phase error rate in the $\mathcal{L}$-bit string, which can be similarly estimated by
\begin{align}\label{n1Lm}
	n^{1}_{\mathcal{L}} \geq \mathcal{L}\left[\frac{n^\prime_{1}}{n^\prime_{Z}} - \gamma^U\left(\mathcal{L}, n^\prime_{Z} - \mathcal{L}, \frac{n^\prime_{1}}{n^\prime_{Z}}, \epsilon_p\right)\right],  
\end{align}
\begin{align}\label{ephLm}
	e^{ph}_{\mathcal{L}} \leq e_{ph}^\prime + \gamma^U(n^{1}_{\mathcal{L}}, n^\prime_{1} - n^{1}_{\mathcal{L}}, e_{ph}^\prime, \epsilon_p),
\end{align}
Note that in multi-bit scheme, the error correction are required, hence the min-entropy is 
\begin{align}
	\mathcal{H}_\mathcal{L}  \leq n^{1}_{\mathcal{L}} [1 - H(e^{ph}_{\mathcal{L}})] - \lambda_{EC},\label{Hnm}
\end{align}
and $\lambda_{EC} = \mathcal{L} fH(E^\prime_Z)$ is the information consumed in the error correction process of the $\mathcal{L} $-bit string.

In Eqs. (\ref{n1Ls}), (\ref{ephLs}), (\ref{n1Lm}) and (\ref{ephLm}), the function $\gamma^U(n,k,\lambda,\epsilon_p)$ are given by
\begin{equation}
	\gamma^U(n,k,\lambda,\epsilon_p) = \frac{\frac{(1-2\lambda)AG}{n+k} + \sqrt{\frac{A^2G^2}{(n+k)^2} + 4\lambda(1-\lambda)G}}{2+2\frac{A^2G}{(n+k)^2}}, 
\end{equation}
in which 
\begin{equation}
	A = \text{max}\{n, k\},
\end{equation}
and
\begin{equation}
	G = \frac{n+k}{nk} \ln \left( \frac{n+k}{2\pi nk\lambda (1-\lambda)\epsilon_p^2} \right).
\end{equation}

\section{SECURITY ANALYSIS}\label{APP:security}
 \subsection {Security of single-bit scheme}
 In this section, we present the security of the single-bit TF-QDS protocol \cite{TFQDS}.
 
 \textbf{\textit{Robustness:}} Robustness indicates the probability of the QDS aborting when Alice, Bob, and Charlie are all honest, which is caused by an error test failure. Through the error rate of test keys, we can estimate the error rate of kept signature keys with a failure probability $\epsilon_{PE}$.
 Considering there are failure possibilities for both processes (Alice-Bob and Alice-Charlie), robustness probability can be expressed as
 \begin{align}\label{PRobust}
 	\epsilon_{rob} = 2\epsilon_{PE}.
 \end{align}
 
 \textbf{\textit{Repudiation:}} The repudiation probability characterizes Alice's signature accepted by Bob but rejected by Charlie. To repudiate, Alice must make the mismatch rate between both elements of $S_m^B$ and the signature $(m,Sig_m)$ lower than $s_a$, and meantime the mismatch rate between either element of $S_m^C$ and the signature $(m,Sig_m)$ to be higher than $s_v$ after the key exchange. Therefore, in this case, the repudiation probability is bounded by
 \begin{align}\label{PRep}
 	\epsilon_{rep} = 2{e^{ - \frac{1}{4}{{\left( {{s_v} - {s_a}} \right)}^2}\mathcal{L}}},
 \end{align}
 where we set ${s_a} = {E^\prime_{Z}} + {({P_e} - {E^\prime_{Z}})}/3 $ and ${s_v} = { E^\prime_{Z}} + 2{({P_e} - {E^\prime_{Z}})}/3$. Note that Eq. (\ref{PRep}) decays exponentially as $\mathcal{L}$ increases, hence we can also choose a larger signature length $\mathcal{L}$ to make the repudiation probability small enough as we expect.
 
 \textbf{\textit{Forgery:}} The forging indicates that the signature is not signed by Alice but would be accepted by Bob and Charlie. In order to forge, the forger must keep the mismatch rate between his declaration $(m,Sig_m)$ and Bob's keys $S_m^B$ and Charlie's keys $S_m^C$ being lower than given values $s_a$ and $s_v$. Bob is the most convenient forger, hence the probability of Bob making fewer than $s_v \mathcal{L}/2$ errors in his declared signature is restricted except with probability at most
 \begin{align}\label{epsilon}
 	{\epsilon_F} = \frac{1}{g}\left( {{2^{ - [ { \mathcal{H}_\mathcal{L} - \frac{\mathcal{L}}{2}{H_2}({s_v})} ]}} + \epsilon_p} \right).
 \end{align}
 The forging probability includes all the process guessing the Charlie's kept keys, which is a composable security given by
 \begin{align}\label{PForge}
 	{\rm{P(Forge)}} \leqslant g + \epsilon_{F} + \epsilon_{PE} + \epsilon_{n^1_{\mathcal{L}}} + \epsilon_{e^{ph}_{\mathcal{L}}}.
 \end{align}
 where $\epsilon_{n^1_{\mathcal{L}}}$ and $\epsilon_{e^{ph}_{\mathcal{L}}}$ are the error probabilities related to the estimation of $n^1_{\mathcal{L}}$, and $e^{ph}_{\mathcal{L}}$, respectively. With the procedure of estimating $n^1_{\mathcal{L}}$ and $e^{ph}_{\mathcal{L}}$, we can count $\epsilon_{n^1_{\mathcal{L}}}=11\epsilon_{p}$ and $\epsilon_{e^{ph}_{\mathcal{L}}}=20\epsilon_{p}$.

 \subsection {Security of multi-bit scheme}
 In this section, we present the security of the TF-QDS protocol with the OTUH method \cite{Li2023}.
 
 \textbf{\textit{Robustness:}} The protocol aborts in an honest execution only if Alice and Bob (or Charlie) disagree on their final keys after the distribution stage. Since error correction ensures key consistency, abortion occurs solely due to transmission errors. The robustness error is bounded by
 \begin{align}
 	\epsilon_{rob} = 2\epsilon_{cor} + 2\epsilon',
 \end{align}
 where $\epsilon_{cor}$ is the error correction failure probability, and $\epsilon'$accounts for classical channel errors.
 
 \textbf{\textit{Repudiation:}} Repudiation succeeds if Bob accepts a message while Charlie rejects it. However, when both parties are honest and share identical keys, their decisions will always match. Thus, repudiation can only occur if errors disrupt the key exchange. The repudiation bound is 
 \begin{align}
 \epsilon_{rep} = 2\epsilon'.
\end{align}
 
 \textbf{\textit{Forgery:}} A forgery succeeds when Charlie accepts a message tampered by Bob. According to the our TF-QDS protocol, Charlie accepts a message only if both the one-time pad decryption and AXU hash verification pass. This scenario is fundamentally equivalent to the authentication attack model described in \cite{Li2023}, where Bob acts as the adversary attempting to forge messages from Alice to Charlie. The forgery success probability is
 \begin{align}
 \epsilon_{for} = m\cdot2^{-2-\mathcal{H}_\mathcal{L} }.
\end{align}

 Finally, the overall security bound of the protocol, defined as the maximum failure probability $\epsilon$, 
 
 \begin{align}
 	\epsilon &\geq max\{\epsilon_{rob}, \epsilon_{rep}, \epsilon_{for}\}.
 \end{align}
 Note that in above security analysis, the security parameters including $\epsilon_{p}$, $\varepsilon$, $\epsilon_{PE}$, $g$, $\epsilon_{cor}$, $\epsilon'$ are set as $10^{-12}$.

\section{EXPERIMENTAL DATA AND RESULTS}\label{APP:data}
The experimental parameters of setup, observed values, estimated parameters and corresponding results are summarized in Table. \ref{Experiment:data}. Note that in single-bit scheme and multi-bit scheme, the corresponding keys length for signature are $\mathcal{L}_1$ and $\mathcal{L}_2$, respectively, which are individually optimized for each scheme and distance to obtain the maximal signature rate. 
\begin{table*}[htbp]
	\centering
	\caption{Experimental data and results. Fiber Length: the total fiber length between Alice and Bob (Charlie); Loss: the total loss of the fiber between Alice and Bob(Charlie); $F$: the effective system frequency; $N_{tot}$: the total number of signal pulses; $u$ ($v$): the intensity of the signal (decoy) pulse; $p_v$ ($p_o$): the sending probability of the decoy (vacuum) pulse; $n_{xy}$: the number of detection events of Alice sending source $x$ and Bob (Charlie) sending source $y$, where $x, y \in \{o,v,u\}$; $m_{vv}$: the number of detection events of Alice sending source $v$ and Bob (Charlie) sending source $v$ in same and anti phase slices; $n_Z$, $E_Z$, $n_1$ and $e_{ph}$ ($n^\prime_Z$, $E^\prime_Z$, $n^\prime_1$ and $e^\prime_{ph}$) are the raw key bits, error rate, estimated single-photon counts and phase error rate in $Z$ windows before AOPP (after AOPP), respectively; $\mathcal{L} _1$, $n^1_{\mathcal{L} _1}$, $e^{ph}_{\mathcal{L} _1}$ and $R_{single}$ ($\mathcal{L} _2$, $n^1_{\mathcal{L} _2}$, $e^{ph}_{\mathcal{L} _2}$ and $R_{multi}$) are the bits length for signature,  estimated single-photon counts in $\mathcal{L} _1$ ($\mathcal{L} _2$) bits, estimated single-photon phase error rate in $\mathcal{L} _1$ ($\mathcal{L} _2$) bits, signature rate of single-bit (multi-bit) scheme, respectively.	}
	\begin{ruledtabular}
		\begin{tabular}{@{}ccccc@{}}
			Fiber Length & \multicolumn{2}{c}{302 km} & \multicolumn{2}{c}{504 km}  \\
			\cline{2-3} \cline{4-5}
			Link  & Alice-Bob & Alice-Charlie & Alice-Bob & Alice-Charlie  \\
			
			Loss (dB) & 57.4 & 57.4 & 95.8 & 95.8 \\
			
			F (GHz) & 1 & 1 & 1 & 1 \\
			
			$u$  & 0.418 & 0.418 & 0.424 & 0.424 \\
			
			$v$  & 0.0264 & 0.0264 & 0.1 & 0.1 \\
			
			$p_o$  & 0.0093 & 0.0093 & 0.024 & 0.024 \\
			
			$p_v$  & 0.0197 & 0.0197 & 0.049 & 0.049 \\
			
			$p_s$  & 0.28 & 0.28 & 0.28 & 0.28 \\
			
			$N$  & $10^{14}$ & $10^{14}$ & $10^{14}$ & $10^{14}$ \\
			\hline
			$n_{oo}$  & 7 & 5 & 233 & 56 \\
			
			$n_{ov}$  & 397043 & 393944 & 151667 & 136596 \\
			
			$n_{vo}$  & 385695 & 370335 & 144307 & 153476 \\
			
			$n_{ou}$  & 84552427 & 82983350 & 3340314 & 3320024 \\
			
			$n_{uo}$  & 83052633 & 84535445 & 3296684 & 3273380 \\
			
			$m_{vv}$  & 3822 & 3948 & 1490 & 1389 \\
			
			$n_Z$ (Before AOPP) & 17353767556 & 17361970696 & 247877706 & 255294247 \\
			
			$n^\prime_{Z}$ (After AOPP)  & 3514754061 & 3507806729 & 5.0279301 & 51913395 \\
			
			$E_Z$ (Before AOPP) & 28.2\% & 28.1\% & 28.5\%  & 27.9\%  \\
			
			$E^\prime_Z$ (After AOPP)  & $5.71\times10^{-6}$ & $2.00\times10^{-5}$ & $1.21\times10^{-3}$  & $3.14\times10^{-4}$  \\
			
			$n_1$ (Before AOPP) & 8315412214 & 8098661172 & 113656381  & 111062741  \\
			
			$n^\prime_1$ (After AOPP)  & 1561633524 & 1473015353 & 20528600  & 18630179  \\
			
			$e_{ph}$ (Before AOPP)  & 2.23\% & 2.36\% & 3.03\%  & 2.96\%  \\
			
			$e^\prime_{ph}$ (After AOPP)   & 4.37\% & 4.62\% & 5.99\%  & 5.91\%  \\					
			\hline
			$\mathcal{L} _1$      & \multicolumn{2}{c}{329841} & \multicolumn{2}{c}{689647} \\
			$n^1_{\mathcal{L} _1}$ &\multicolumn{2}{c}{67939}  & \multicolumn{2}{c}{125916}  \\
			$e^{ph}_{\mathcal{L}_1 }$ & \multicolumn{2}{c}{5.19\%}  & \multicolumn{2}{c}{6.46\%}  \\
			$R_{single}$ (bps)  & \multicolumn{2}{c}{0.0532} & \multicolumn{2}{c}{$3.65\times10^{-4}$} \\
%			\cline{1-1} \cline{2-3} \cline{4-5}
			\hline
			$\mathcal{L} _2$      & \multicolumn{2}{c}{832} & \multicolumn{2}{c}{1148} \\
			$n^1_{\mathcal{L} _2}$ &\multicolumn{2}{c}{249}  & \multicolumn{2}{c}{308}  \\
			$e^{ph}_{\mathcal{L} _2}$ &\multicolumn{2}{c}{23.9\%}  & \multicolumn{2}{c}{23.1\%}  \\
			$R_{multi}$ (tps)  & \multicolumn{2}{c}{21.1} & \multicolumn{2}{c}{0.219} \\
		\end{tabular}
	\end{ruledtabular}
	\label{Experiment:data}
\end{table*}

%\bibliography{TFQDS.bib}

\end{document}